\documentclass{mn2e}%

\renewcommand\[{\begin{equation}}
\renewcommand\]{\end{equation}}

\def\d{{\rm d}}
\def\p{\partial}
\def\i{\relax\ifmmode{\rm i}\else\char16\fi}

\def\lesssim{{_ <\atop{^\sim}}}

\catcode`\@=11 
\def\@versim#1#2{\vcenter{\offinterlineskip
        \ialign{$\m@th#1\hfil##\hfil$\crcr#2\crcr\sim\crcr } }}
\def\lsim{\mathrel{\mathpalette\@versim<}}
\def\gsim{\mathrel{\mathpalette\@versim>}}

\def\b#1{{\bf{#1}}}

\def\lesssim{\mathrel{\hbox{\rlap{\hbox{\lower4pt\hbox{$\sim$}}}\hbox{$<$}}}}
\def\gtrsim{\mathrel{\hbox{\rlap{\hbox{\lower4pt\hbox{$\sim$}}}\hbox{$>$}}}}

\def\apj#1 #2{ApJ, #1, #2}
\def\aj#1 #2{AJ, #1, #2}
\def\mn#1 #2{MNRAS, #1, #2}
\def\aa#1 #2{A\&A, #1, #2}

\begin{document}

   \title[Two-temperature Accretion Flows] {On the Viability of Two-temperature Accretion Flows}

   \author[Quataert]
          {Eliot Quataert
           \\Astronomy Department, 601 Campbell Hall, University of California, Berkeley, CA 94720; eliot@astro.berkeley.edu \\
          }

   \date{}

   \maketitle

\begin{abstract}
Binney (2003) has argued that two-temperature radiatively inefficient
accretion flow models are unphysical because the electron-ion
equipartition time is much shorter than the accretion time.  I show
that this conclusion is incorrect because it relies on a
misidentification of the electron-ion equipartition time.  I also
clarify what requirements must, in fact, be satisfied to maintain a
two-temperature accretion flow.

\end{abstract}


\section{Introduction}

Since the mid 1970's accretion theorists have invoked two-temperature
collisionless plasmas in models of accreting compact objects (e.g.,
Shapiro, Lightman, \& Eardley 1976; Ichimaru 1977; Rees et al. 1982;
Narayan \& Yi 1995).  In such models it is usually assumed that
Coulomb collisions are the dominant mechanism for exchanging energy
between electrons and ions.  The inefficiency of collisional energy
transfer at high temperatures and low densities (accretion rates),
together with the much shorter radiative cooling time of relativistic
electrons with respect to nonrelativistic ions, leads to the
possibility of a two-temperature plasma with $T_i \gg
T_e$.\footnote{Even in the absence of radiative cooling, a
two-temperature plasma can develop because the relativistic electrons
have a smaller adiabatic index than the non-relativistic ions.} An
outstanding problem is whether there are any mechanisms that might
enforce electron-ion energy equipartition on a timescale much shorter
than that of Coulomb collisions (e.g., Phinney 1983; Begelman \&
Chiueh 1988).  If so, then two-temperature accretion flow models would
not be viable.

Binney (2003; hereafter B03) has argued that the electron-ion
equipartition time is, in fact, much shorter than the accretion time,
and thus that two-temperature accretion flow models are unphysical.
His argument is actually more general, and implies that electron-ion
equipartition is a characteristic feature of plasmas in the presence
of a time varying electromagnetic field (in the accretion context, MHD
turbulence driven by the magnetorotational instability (MRI; Balbus \&
Hawley 1991) is an important and widely studied example of such a time
varying electromagnetic field).  In the next section I briefly review
Binney's argument and explain why it is incorrect. I then summarize
the conditions that must be satisfied to maintain a two-temperature
accretion flow, and what calculations can be done (and have been done)
to address this issue.
\section{Electron-Ion Equipartition}
B03 estimates the electron-ion equipartition time by first evaluating
the work done on a particle by an electromagnetic field (described by
potentials $\psi$ and $\b A$).  The Hamiltonian for the motion of a
particle of mass $m$ and charge $q$ is
\[\label{H} H={(\b p-q\b A)^2\over2m}+q\psi+m\Phi.
\]  where $\Phi$ is the gravitational potential. The change in the 
particle's energy is given by $\dot H$, i.e.,
\begin{eqnarray}\label{Hdot} {\d H\over\d t}&=&{\p H\over\p t} =-q{\b
p-q\b A\over m}\cdot{\p\b A\over\p t}+q{\p\psi\over\p t}\nonumber\\
&=&-q\b v\cdot{\p\b A\over\p t}+q{\p\psi\over\p t},
\end{eqnarray}
where, as in B03, I have assumed that $\Phi$ is time-independent.
Equation (\ref{Hdot}) is the work done by the electric field.


Using equations (\ref{H}) and (\ref{Hdot}) we can define the timescale
on which the energy of a particle changes as
\[\label{tE} t_{\rm E} \equiv {H\over|\d H/\d
t|}.\] B03 identifies $t_E$ in equation (\ref{tE}) as the timescale
for electrons and ions in a two-temperature plasma to come into
equipartition, mediated by the electromagnetic field.  His argument
for this identification is that ``since the right side of this
equation [eq. (\ref{Hdot})] is proportional to the charge $q$, if at
some location energy is lost by one species, it is gained by the
oppositely charged species. Thus this equation describes the mechanism
by which equipartition is established between ions and electrons; the
net direction of the energy flow is mandated by the general principles
of statistical physics, and the rate of flow may be estimated from
equation (\ref{Hdot}).''

This interpretation of equation (\ref{Hdot}) is incorrect, as is the
resulting identification of equation (\ref{tE}) as the equipartition
time.  The essential problem is that equation (\ref{Hdot}) represents
the instantaneous work done on a particle by the electric field.  It
does not, however, represent the {\it net} energy transfer integrated
over time, i.e., the true heating or change in entropy.  These differ
because of the adiabaticity of particle motion in a plasma and because
energy transfer occurs at discrete resonances (Landau and cyclotron)
that are not accounted for in equation [3] (see below).  

To elaborate on these points, it is useful to focus on a concrete
example:\footnote{Because there is no general proof for or against the
claim that electron-ion thermal equilibration in a collisionless
plasma is rapid, this example is intended only to show why B03's
conclusions are incorrect; the more general problem of how to assess
whether a two-temperature accretion flow can in fact be maintained is
discussed in the next section.} consider a magnetic field that varies
on a timescale $t$ that is much longer than $\Omega^{-1}$, where
$\Omega = eB/mc$ is the cyclotron frequency.  The magnetic moment of a
particle is then an adiabatic invariant, i.e., $\mu = mv_{\perp}^2/B =
{\rm constant}$, where $v_\perp$ is the velocity perpendicular to the
magnetic field (e.g, Sturrock 1994).  Thus, if the magnetic field
varies in an arbitrary (slow) manner, and returns to its initial
value, the perpendicular kinetic energy (temperature) of {each
particle, and thus each particle species}, is ultimately unchanged.
This is in spite of the fact that the energy of every particle
instantaneously changes with the magnetic field on the timescale $\sim
t$.  By contrast, B03's argument, which considers only this
instantaneous change in energy, incorrectly implies that 1.
equipartition is established, and 2.  that it is established on a
timescale given by equation (3).  This example highlights that the
timescale given by equation (3) is in general the timescale for
adiabatic changes, not true heating (let alone equipartition).  A
similar conclusion could be drawn by considering the energy of a
particle in the presence of an undamped Alfv\'en wave: during the
oscillation of the wave magnetic energy gets converted into particle
energy and vice-versa, but there is no net transfer of energy to or
from the particles, and no tendency towards equipartition, during a
period of the wave.

\section{Discussion}

Although the arguments in the previous section show the problems with
Binney's claim that electron-ion equipartition is rapid, they do not
specify how to in fact determine whether a two-temperature accretion
flow can be maintained.  I believe that there are two key calculations
that must be done to address this.  First, given that MHD turbulence
generated by the MRI drives accretion, it is important to understand
how the MRI operates in a collisionless plasma and how the resulting
turbulent energy is transferred to the particles.  There has been a
fair amount of analytical work addressing this (e.g., Bisnovatyi-Kogan
\& Lovelace 1997; Quataert 1998; Gruzinov 1998; Blackman 1999;
Quataert \& Gruzinov 1999; Medvedev 2000; Sharma et al. 2003).  The
key physics is that turbulent energy is transferred to particles at
discrete wave-particle resonances, when $\omega - k_\parallel
v_\parallel \approx n \Omega$, where $n$ is an integer, $v_\parallel$
is the velocity of a particle along the local magnetic field, and
$\omega$ and $k_\parallel$ are the frequency and parallel wavevector
of a wave comprising the turbulence.  The details of which particles
are heated depend on which waves are present in the turbulence and
which particles are resonant; the net particle heating will in general
be different for the ions and the electrons and so there is no reason
for the system to approach equipartition.  Existing calculations of
particle heating using models for MRI-generated turbulence are
somewhat uncertain and depend sensitively on $\beta$ (the ratio of the
gas pressure to magnetic pressure in the flow), with $\beta \sim 1$
favoring electron heating while higher $\beta \gsim 10$ favors ion
heating.  This uncertainty stems from (1) uncertainty in the relative
importance of the three MHD waves in MRI-generated turbulence, (2)
uncertainty in the importance of magnetic reconnection (e.g.,
Bisnovatyi-Kogan \& Lovelace 1997; Blackman 1999), and (3) uncertainty
in the behavior of Alfv\'enic turbulence on small scales comparable to
the proton Larmor radius (see Quataert \& Gruzinov 1999).  Numerical
simulations of turbulence in collisionless plasmas are currently
underway that will help sort out these issues.
 
Even if MRI-generated turbulence predominantly heats the ions, there
is no guarantee that the resulting two-temperature plasma is stable.
There could be kinetic instabilities that transfer ion thermal energy
to the electrons on a timescale much shorter than the accretion time,
thus enforcing a one-temperature plasma.  It is important to stress
that the MRI is not an example of such an instability because it feeds
on the free energy of differential rotation, not the ion thermal
energy.  It is also worth noting that a uniform two-temperature
collisionless plasma containing particles with Maxwellian distribution
functions is stable to linear perturbations (e.g., Stix 1992).  Thus a
candidate instability must feed on gradients in the background medium
or velocity space anisotropies, not the mere presence of a
two-temperature plasma.  The question of whether there are any such
instabilities in the accretion flow context is an open and difficult
problem (see Begelman \& Chiueh 1988 for the most detailed
investigation to date). In the absence of theoretical calculations or
numerical simulations to settle this issue, it may be useful to take
observations as our guide: both the solar wind and the post-shock
environment in supernova remnants show two-temperature plasmas in
which it appears that the dominant mechanism for electron-ion energy
exchange is Coulomb collisions, not a kinetic instability (see, e.g.,
Esser et al. 1999 for the solar wind and Michael et al. 2002 for
supernova remnants).  This qualitatively supports the assumptions of
two-temperature accretion flow models, with the necessary caveat that
the plasma conditions in these examples are quite different from those
around compact objects.

\section*{acknowledgments}  I thank my colleagues for their many emails 
which prodded me to write this short paper.  Thanks as well to Jon
Arons, Bill Dorland, Greg Hammett, and Ramesh Narayan for useful
conversations and advice.

\end{document}